# Dual-horizon peridynamics: A stable solution to varying horizons

Huilong Ren[c], Xiaoying Zhuang[d,e], Timon Rabczuk[a,b,c,*]

[a] *Division of Computational Mechanics, Ton Duc Thang University, Ho Chi Minh City, Viet Nam*
[b] *Faculty of Civil Engineering, Ton Duc Thang University, Ho Chi Minh City, Viet Nam*
[c] *Institute of Structural Mechanics, Bauhaus-Universität Weimar, 99423 Weimar, Germany*
[d] *State Key Laboratory of Disaster Reduction in Civil Engineering, College of Civil Engineering, Tongji University, Shanghai 200092, China*
[e] *Institute of Continuum Mechanics, Leibniz Universität Hannover, Hannover, Germany*

## Highlights

- Stable crack paths with variable horizon and particle sizes.
- Dual-horizon peridynamics for multiple materials/material interfaces.
- Fulfillment of balance of momentum and balance of angular momentum.

## Abstract

In this paper, we present a dual-horizon peridynamics formulation which allows for simulations with dual-horizon with minimal spurious wave reflection. We prove the general dual property for dual-horizon peridynamics, based on which the balance of momentum and angular momentum in PD are naturally satisfied. We also analyze the crack pattern of random point distribution and the multiple materials issue in peridynamics. For selected benchmark problems, we show that DH-PD is less sensitive to the spatial than the original PD formulation.
ⓒ 2016 Elsevier B.V. All rights reserved.

## 1. Introduction

Peridynamics (PD) has recently attracted broad interests for researchers in computational solid mechanics since it shows great flexibility in modeling dynamic fractures. The crack happens naturally, and no attention is paid to describe crack topology in the PD simulation. Hence, complicated fracture patterns i.e. crack branching and coalescence of multiple cracks, can be captured with ease by breaking the bonds between material points. No techniques such as smoothing the normals of the crack surfaces in order to avoid erratic crack paths is necessary in PD, while such techniques are commonly used in the extended finite element method (XFEM) [1], meshless methods [2] or other partition of unity methods (PUM) [3]. Meanwhile, PD can be easily extended from 2D to 3D, which cannot always be done in other methods [4–6]. In addition, an important feature in PD is that it can now be easily coupled with the more classical local continuum mechanics [7,8]. The traditional PD method was proposed by Silling [9] and has been applied onwards for a wide range of problems including impact loading, fragmentation [10,11], composites delamination [12], beam and plate/shell structures [13–15], thermal diffusion [16] and flow in porous media [17,18].

---

* Corresponding author at: Division of Computational Mechanics, Ton Duc Thang University, Ho Chi Minh City, Viet Nam.
  *E-mail addresses:* zhuang@ikm.uni-hannover.de (X. Zhuang), timon.rabczuk@tdt.edu.vn (T. Rabczuk).

In PD, the equation of motion is formulated in an integral form rather than the partial differential form. The material point interacts with other material points in its horizon, a domain associated to the point. The horizon is often represented by circular area (2D) or spherical region (3D) with a radius. There are mainly 3 types of PD formulations, i.e. bond based peridynamics (BB-PD), ordinary state based peridynamics (OSB-PD) and non-ordinary state based peridynamics (NOSB-PD). PD formulation starts with discretizing the solid domain into material points, where any two material points within horizon's radius form a bond. The first version of PD is called the bond based formulation (BB-PD) [9] as the bonds act like independent springs. One limitation of BB-PD is that only the material with Poisson's ratio of 1/3 in 2D or 1/4 in 3D can be modeled. In order to eliminate the Poisson's ratio restriction, the state based peridynamics (SB-PD) [19] is proposed, where "state" means the bond deformation depends not only on the deformation of the bond itself, but also on collective deformation of other bonds. The SB-PD was extended into two types, namely the OSB-PD and the NOSB-PD. After that, BB-PD is regarded as a special case of OSB-PD. NOSB-PD is capable of simulating the material with advanced constitute models, e.g. viscoplasticity model [20], the Drucker–Prager plasticity model [10] and Johnson–Cook (JC) constitutive model [21].

Aside from the pure PD formulation above mentioned, much work was devoted to the coupling of peridynamics with the continuum methods [7,8,22–24]. Such strategies enable to model the fracture domain with PD while other domain modeled with continuum methods. Seleson et al. developed a force-based coupling scheme for peridynamics and classical elasticity in one dimension [8] and higher dimensions [22]. This method satisfies the Newton's third law and Patch test, and no ghost force exists in the blending domain. Lubineau et al. developed a morphing strategy to couple non-local to local continuum mechanics. The strategy is based on the domain transition where the strain energy is divided into nonlocal part and local continuum part. In this method, the PD is discretized with finite element and the local integration matrix is assembled.

One basic constraint of traditional PD formulations is that the radii of horizons are required to be constant, otherwise spurious wave reflections shall emerge and ghost forces between material points will deteriorate the results. However, in many applications, for sake of computational efficiency, it is necessary to vary the horizon sizes with respect to the spatial distribution of the material points, e.g. for adaptive refinement, and multiscale modeling. In the implementation of traditional PD discretized with particles, in order to achieve acceptable accuracy, the horizon radius has to be determined with respect to the lowest material point resolution locally required. It should be noted that horizon is guided by the characteristic range of long-range forces [9,7], not the discretization of the computational domain.

The dual-horizon peridynamics (DH-PD) is firstly proposed to solve the issue of spurious wave reflections when variable horizons are adopted. The key idea lies in the definition of dual-horizon, which is defined as the dual term of the horizons centering at each material points. Within the framework of DH-PD, the traditional peridynamics with constant horizon can be derived. The derivation of DH-PD is based on the Newton's third law and other additional techniques such as variational principal or Taylor expansion [25] are not required. The dual-horizon PD not only considers the variable horizons, but also reduce the computational cost, allowing the domain discretized like finite element method(FEM) or finite volume method (FVM), where the domain of interest utilizes dense node distribution and the other domain coarse node distribution.

In this paper, we review the DH-PD and propose a universal principle in dual-horizon, the dual property of dual-horizon. We also discuss the influence of random material points distribution on crack pattern and multiple materials issue in peridynamics. The content of the paper is outlined as follows. Section 2 begins with the equation of motion of dual-horizon peridynamics and then discuss three ways to calculate the bond force in DH-PD. Section 3 proves the general dual property of dual-horizon, based on which the balance of linear momentum and the balance of angular momentum of the present PD formulation are re-proved. In Section 4, the issue of multiple materials in peridynamics is discussed with 2 numerical examples. In Section 5, the simulation of Kalthoff–Winkler experiment is carried out to test the influence of random material points distribution on the crack pattern in peridynamics.

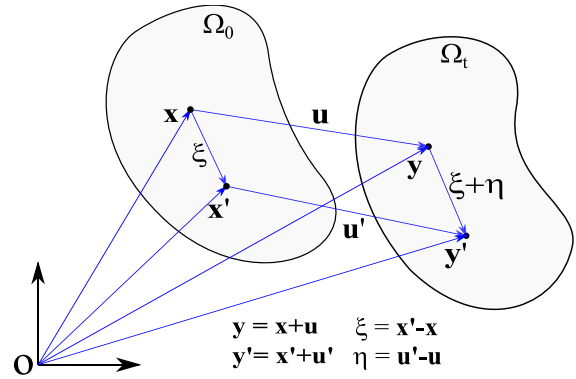

Fig. 1. The configuration for deformed body.

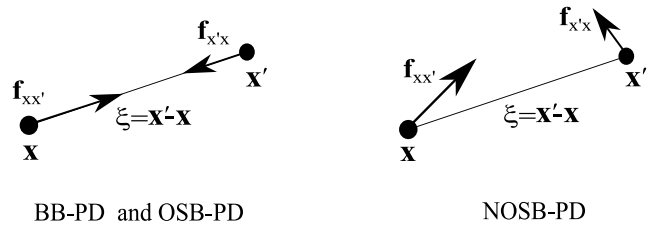

Fig. 2. Force vectors of BB-PD, OSB-PD and NOSB-PD.

## 2. Dual-horizon peridynamics

Consider a solid domain in the initial and current configuration as shown in Fig. 1. $\mathbf{x}$ is the material coordinate in initial configuration $\Omega_0$; and $\mathbf{y} := \mathbf{y}(\mathbf{x}, t)$ and $\mathbf{y}' := \mathbf{y}(\mathbf{x}', t)$ are the spatial coordinates in the current configuration $\Omega_t$; $\mathbf{u} := \mathbf{u}(\mathbf{x}, t)$ and $\mathbf{u}' := \mathbf{u}(\mathbf{x}', t)$ are the displacement vectors for associated points; $\mathbf{v} := \mathbf{v}(\mathbf{x}, t)$ and $\mathbf{v}' := \mathbf{v}(\mathbf{x}', t)$ are the velocity for associated points; $\boldsymbol{\xi} := \mathbf{x}' - \mathbf{x}$ denotes the initial bond vector; the relative displacement vector of bond $\boldsymbol{\xi}$ is $\boldsymbol{\eta} := \mathbf{u}' - \mathbf{u}$, and the current bond vector is $\mathbf{y}\langle\boldsymbol{\xi}\rangle := \mathbf{y}(\mathbf{x}', t) - \mathbf{y}(\mathbf{x}, t) = \boldsymbol{\xi} + \boldsymbol{\eta}$.

Since the bond $\boldsymbol{\xi}$ starts from $\mathbf{x}$ to $\mathbf{x}'$, an alias $\mathbf{xx}'$ is used to denote $\boldsymbol{\xi}$. Obviously, $\mathbf{x}'\mathbf{x}$ is in opposite direction of $\mathbf{xx}'$. Let $\mathbf{f}_{\mathbf{xx}'} := \mathbf{f}_{\mathbf{xx}'}(\boldsymbol{\eta}, \boldsymbol{\xi})$ denote the force vector density acting on point $\mathbf{x}$ due to bond $\mathbf{xx}'$; based on Newton's third law, $\mathbf{x}'$ undertakes a reaction force $-\mathbf{f}_{\mathbf{xx}'}$.

As the conventional peridynamics formulation with constant horizon can be viewed as a special case of the DH-PD [26], we here mainly focus on the DH-PD. There are general three types of peridynamics, as shown in Fig. 2 and DH-PD will be applied to these three types.

### 2.1. The shortcomings of constant horizons

In conventional peridynamics, the horizon is required to be constant for a homogeneous body, where being homogeneous means that the strain energy density for any uniform deformation must be invariant with respect to changes in $\delta$ [27]. When the computational domain is discretized with different material point sizes, as shown in Fig. 3, in order to avoid the spurious wave reflection, the number of neighbors in the right side is dramatically increased. The fact of too many neighbors not only increases the computational cost, but also contributes little to the accuracy improvement, as the value in the point is smoothed out by its neighbors.

When different horizon radii are used, spurious wave reflections occur in conventional peridynamics. The reason is that the traditional peridynamics formulation with single horizon is not easy to account for the interaction between two points with different horizon radii, which has been discussed in detail in [26].

### 2.2. Horizon and dual-horizon

In conventional PD, the *horizon* is a domain centered at each point with a radius of $\delta$. In DH-PD formulation, the *horizon* $H_\mathbf{x}$ is the domain where any material point $\mathbf{x}'$ falling inside forms bond $\mathbf{xx}'$, where the bond $\mathbf{xx}'$ will exert direct force $\mathbf{f}_{\mathbf{xx}'}$ on $\mathbf{x}$ and whereas $\mathbf{x}'$ will undertake reaction force $-\mathbf{f}_{\mathbf{xx}'}$. In this sense, the horizon in DH-PD can be viewed as direct force horizon. The *dual-horizon* in DH-PD is defined as a union of points whose horizons include $\mathbf{x}$, denoted by

$$H'_\mathbf{x} = \{\mathbf{x}' | \mathbf{x} \in H_{\mathbf{x}'}\}. \tag{1}$$

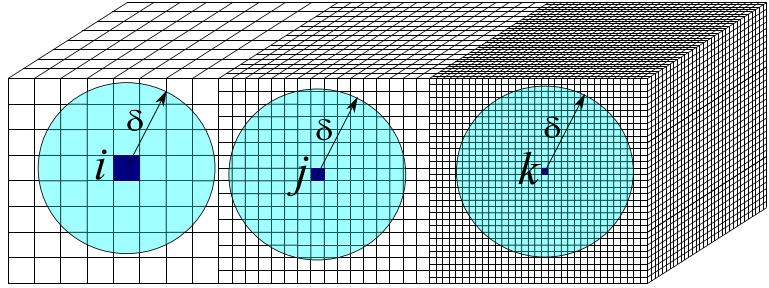

Fig. 3. The neighbors for points $i$, $j$ and $k$ with constant horizon radius $\delta = 3r_i$. $r_i = 2r_j = 4r_k$, neighbor number $N_i \approx 200$, $N_j \approx 2^3 N_i \approx 1600$, $N_k \approx 4^3 N_i \approx 12{,}800$.

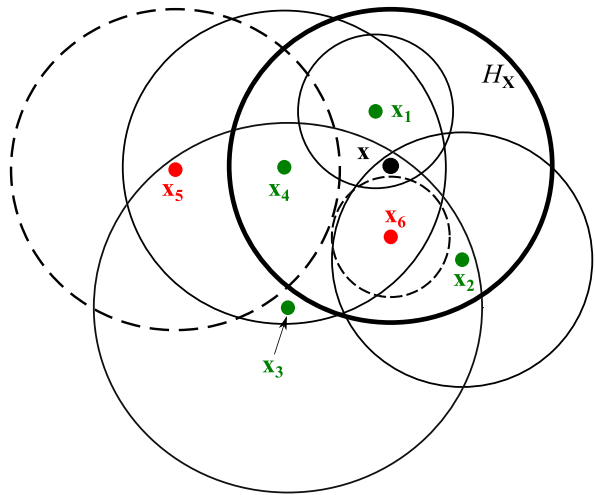

Fig. 4. The schematic diagram for horizon and dual-horizon, all circles above are horizons. The green points $\{\mathbf{x}_1, \mathbf{x}_2, \mathbf{x}_3, \mathbf{x}_4\} \in H'_\mathbf{x}$, whose horizons are denoted by thin solid line; the red points $\{\mathbf{x}_5, \mathbf{x}_6\} \notin H'_\mathbf{x}$, whose horizons are denoted by dashed line. (For interpretation of the references to color in this figure legend, the reader is referred to the web version of this article.)

The bond from the dual-horizon is named as dual-bond. For any point $\mathbf{x}' \in H'_\mathbf{x}$, the dual-bond $\mathbf{x}'\mathbf{x}$ is actually the bond $\mathbf{x}'\mathbf{x}$ from $H_{\mathbf{x}'}$. Dual-bond $\mathbf{x}'\mathbf{x}$ exerts reaction force on $\mathbf{x}$. In this sense, the dual-horizon can be viewed as reaction force horizon. One simple example to illustrate the dual-horizon is shown in Fig. 4, where $\{\mathbf{x}_5, \mathbf{x}_6\} \notin H'_\mathbf{x}$ because $H_{\mathbf{x}_5}$ and $H_{\mathbf{x}_6}$ do not contain $\mathbf{x}$. The direct force and reaction force from horizon and dual-horizon are shown in Fig. 5.

Bond and dual-bond are broken independently in the fracture models. There are reasons accounting for it. For example, in BB-PD, the equivalence of energy release rate of homogeneous body in fracture models indicates that different horizon sizes correspond to different critical stretches. This means bond $\mathbf{xx}'$ and dual-bond $\mathbf{x}'\mathbf{x}$ may have different critical stretches. If they are broken at the same bond stretch, the one with larger critical bond stretch is broken in advance, which indicates a lower material strength for that point and thus violates the equivalence of energy release rate.

### 2.3. Equation of motion for dual-horizon peridynamics

The internal forces that are exerted at each material point include two parts, the direct forces from the horizon and the reaction forces from the dual-horizon. The other forces applied to a material point include the body force and the inertia force. Let $\Delta V_\mathbf{x}$ denote the volume associated with $\mathbf{x}$. The body force for $\mathbf{x}$ can be expressed as $\mathbf{b}(\mathbf{x}, t)\Delta V_\mathbf{x}$, where $\mathbf{b}(\mathbf{x}, t)$ is the body force density. The inertia force is denoted by $\rho\ddot{\mathbf{u}}(\mathbf{x}, t)\Delta V_\mathbf{x}$, where $\rho$ is the density.

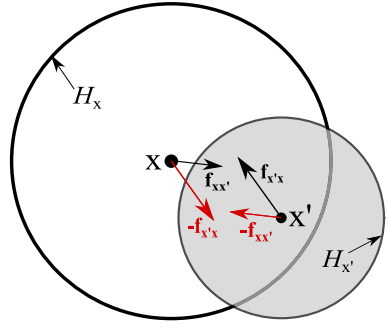

Fig. 5. The force vectors in dual-horizon peridynamics. Bond $\mathbf{xx'}$ exerts direct force $\mathbf{f_{xx'}}$ on $\mathbf{x}$ and reaction force $-\mathbf{f_{xx'}}$ on $\mathbf{x'}$. In this case, $\mathbf{f_{xx'}} \neq 0, \mathbf{f_{x'x}} = 0$.

The direct force due to bond $\mathbf{xx'}$ in $H_\mathbf{x}$ is

$$\mathbf{f_{xx'}}(\boldsymbol{\eta}, \boldsymbol{\xi}) \cdot \Delta V_\mathbf{x} \cdot \Delta V_\mathbf{x'}, \forall \mathbf{x'} \in H_\mathbf{x}. \tag{2}$$

The reaction force due to dual-bond $\mathbf{x'x}$ in $H'_\mathbf{x}$ is

$$-\mathbf{f_{x'x}}(-\boldsymbol{\eta}, -\boldsymbol{\xi}) \cdot \Delta V_\mathbf{x'} \cdot \Delta V_\mathbf{x}, \forall \mathbf{x'} \in H'_\mathbf{x}. \tag{3}$$

By summing over all forces on point $\mathbf{x}$, including inertia force, body force, direct force in Eq. (2) and reaction force in Eq. (3), the equation of motion for DH-PD is obtained.

$$\rho \ddot{\mathbf{u}}(\mathbf{x}, t) \Delta V_\mathbf{x} = \sum_{H_\mathbf{x}} \mathbf{f_{xx'}}(\boldsymbol{\eta}, \boldsymbol{\xi}) \Delta V_\mathbf{x'} \Delta V_\mathbf{x} + \sum_{H'_\mathbf{x}} (-\mathbf{f_{x'x}}(-\boldsymbol{\eta}, -\boldsymbol{\xi}) \Delta V_\mathbf{x'} \Delta V_\mathbf{x}) + \mathbf{b}(\mathbf{x}, t) \Delta V_\mathbf{x}. \tag{4}$$

As the volume $\Delta V_\mathbf{x}$ associated to $\mathbf{x}$ is independent of the summation, we can eliminate $\Delta V_\mathbf{x}$ in Eq. (4), yielding the governing equation based on $\mathbf{x}$:

$$\rho \ddot{\mathbf{u}}(\mathbf{x}, t) = \sum_{H_\mathbf{x}} \mathbf{f_{xx'}}(\boldsymbol{\eta}, \boldsymbol{\xi}) \Delta V_\mathbf{x'} - \sum_{H'_\mathbf{x}} \mathbf{f_{x'x}}(-\boldsymbol{\eta}, -\boldsymbol{\xi}) \Delta V_\mathbf{x'} + \mathbf{b}(\mathbf{x}, t). \tag{5}$$

When the discretization is sufficiently fine, the discrete form converges to the integral form:

$$\lim_{\Delta V_\mathbf{x'} \to 0} \sum_{H_\mathbf{x}} \mathbf{f_{xx'}}(\boldsymbol{\eta}, \boldsymbol{\xi}) \Delta V_\mathbf{x'} = \int_{H_\mathbf{x}} \mathbf{f_{xx'}}(\boldsymbol{\eta}, \boldsymbol{\xi}) \, dV_\mathbf{x'} \tag{6}$$

and

$$\lim_{\Delta V_\mathbf{x'} \to 0} \sum_{H'_\mathbf{x}} \mathbf{f_{x'x}}(-\boldsymbol{\eta}, -\boldsymbol{\xi}) \Delta V_\mathbf{x'} = \int_{H'_\mathbf{x}} \mathbf{f_{x'x}}(-\boldsymbol{\eta}, -\boldsymbol{\xi}) \, dV_\mathbf{x'}. \tag{7}$$

Thus, the integral form of the equation of motion in dual-horizon peridynamics (DH-PD) is given as

$$\rho \ddot{\mathbf{u}}(\mathbf{x}, t) = \int_{H_\mathbf{x}} \mathbf{f_{xx'}}(\boldsymbol{\eta}, \boldsymbol{\xi}) \, dV_\mathbf{x'} - \int_{H'_\mathbf{x}} \mathbf{f_{x'x}}(-\boldsymbol{\eta}, -\boldsymbol{\xi}) \, dV_\mathbf{x'} + \mathbf{b}(\mathbf{x}, t). \tag{8}$$

Note that Eq. (8) is also valid for points near the boundary area.

When the horizons are set constant, as $\mathbf{x'} \in H'_\mathbf{x} \Leftrightarrow \mathbf{x'} \in H_\mathbf{x}$, we have $H'_\mathbf{x} = H_\mathbf{x}$. Then Eq. (8) degenerates to traditional constant horizon peridynamics.

$$\rho \ddot{\mathbf{u}}(\mathbf{x}, t) = \int_{H_\mathbf{x}} \mathbf{f_{xx'}}(\boldsymbol{\eta}, \boldsymbol{\xi}) - \mathbf{f_{x'x}}(-\boldsymbol{\eta}, -\boldsymbol{\xi}) \, dV_\mathbf{x'} + \mathbf{b}(\mathbf{x}, t). \tag{9}$$

### 2.4. Numerical implementation of DH-PD

We aim at solving Eq. (5), the discrete form of DH-PD with the Velocity-Verlet algorithm [28]. After discretization of the domain, each material point's neighbors based on its horizon radius are stored explicitly, while the neighbors in dual-horizon are inferred from other material points' horizons. Every time the bond force from one point's horizon is calculated, the dual-bond force from the other point's dual-horizon is determined according to Newton's third law. For example, assuming the material points with horizons are shown in Fig. 4, the horizons of $\mathbf{x}$ is $H_\mathbf{x} = \{\mathbf{x}_1, \mathbf{x}_2, \mathbf{x}_4, \mathbf{x}_6\}$, and $H'_\mathbf{x} = \{\mathbf{x}_1, \mathbf{x}_2, \mathbf{x}_3, \mathbf{x}_4\}$. Direct forces on $\mathbf{x}$ from its horizon $H_\mathbf{x}$ are summed as follows:

$\mathbf{x}_1 \in H_\mathbf{x}$, calculate $\mathbf{f}_{\mathbf{x}\mathbf{x}_1}$, add $\mathbf{f}_{\mathbf{x}\mathbf{x}_1} \Delta V_{\mathbf{x}_1} \Delta V_\mathbf{x}$ to $\mathbf{x}$, add $-\mathbf{f}_{\mathbf{x}\mathbf{x}_1} \Delta V_{\mathbf{x}_1} \Delta V_\mathbf{x}$ to $\mathbf{x}_1$;

$\mathbf{x}_2 \in H_\mathbf{x}$, calculate $\mathbf{f}_{\mathbf{x}\mathbf{x}_2}$, add $\mathbf{f}_{\mathbf{x}\mathbf{x}_2} \Delta V_{\mathbf{x}_2} \Delta V_\mathbf{x}$ to $\mathbf{x}$, add $-\mathbf{f}_{\mathbf{x}\mathbf{x}_2} \Delta V_{\mathbf{x}_2} \Delta V_\mathbf{x}$ to $\mathbf{x}_2$;

$\mathbf{x}_4 \in H_\mathbf{x}$, calculate $\mathbf{f}_{\mathbf{x}\mathbf{x}_4}$, add $\mathbf{f}_{\mathbf{x}\mathbf{x}_4} \Delta V_{\mathbf{x}_4} \Delta V_\mathbf{x}$ to $\mathbf{x}$, add $-\mathbf{f}_{\mathbf{x}\mathbf{x}_4} \Delta V_{\mathbf{x}_4} \Delta V_\mathbf{x}$ to $\mathbf{x}_4$;

$\mathbf{x}_6 \in H_\mathbf{x}$, calculate $\mathbf{f}_{\mathbf{x}\mathbf{x}_6}$, add $\mathbf{f}_{\mathbf{x}\mathbf{x}_6} \Delta V_{\mathbf{x}_6} \Delta V_\mathbf{x}$ to $\mathbf{x}$, add $-\mathbf{f}_{\mathbf{x}\mathbf{x}_6} \Delta V_{\mathbf{x}_6} \Delta V_\mathbf{x}$ to $\mathbf{x}_6$.

We do not calculate $\mathbf{f}_{\mathbf{x}_1\mathbf{x}}, \mathbf{f}_{\mathbf{x}_2\mathbf{x}}, \mathbf{f}_{\mathbf{x}_3\mathbf{x}}, \mathbf{f}_{\mathbf{x}_4\mathbf{x}}$ from dual-horizon $H'_\mathbf{x}$. The force from $H'_\mathbf{x}$ are determined when calculating forces in $H_{\mathbf{x}_1}, H_{\mathbf{x}_2}, H_{\mathbf{x}_3}, H_{\mathbf{x}_4}$.

$\mathbf{x} \in H_{\mathbf{x}_1}$, calculate $\mathbf{f}_{\mathbf{x}_1\mathbf{x}}$, add $\mathbf{f}_{\mathbf{x}_1\mathbf{x}} \Delta V_\mathbf{x} \Delta V_{\mathbf{x}_1}$ to $\mathbf{x}_1$, add $-\mathbf{f}_{\mathbf{x}_1\mathbf{x}} \Delta V_\mathbf{x} \Delta V_{\mathbf{x}_1}$ to $\mathbf{x}$;

$\mathbf{x} \in H_{\mathbf{x}_2}$, calculate $\mathbf{f}_{\mathbf{x}_2\mathbf{x}}$, add $\mathbf{f}_{\mathbf{x}_2\mathbf{x}} \Delta V_\mathbf{x} \Delta V_{\mathbf{x}_2}$ to $\mathbf{x}_2$, add $-\mathbf{f}_{\mathbf{x}_2\mathbf{x}} \Delta V_\mathbf{x} \Delta V_{\mathbf{x}_2}$ to $\mathbf{x}$;

$\mathbf{x} \in H_{\mathbf{x}_3}$, calculate $\mathbf{f}_{\mathbf{x}_3\mathbf{x}}$, add $\mathbf{f}_{\mathbf{x}_3\mathbf{x}} \Delta V_\mathbf{x} \Delta V_{\mathbf{x}_3}$ to $\mathbf{x}_3$, add $-\mathbf{f}_{\mathbf{x}_3\mathbf{x}} \Delta V_\mathbf{x} \Delta V_{\mathbf{x}_3}$ to $\mathbf{x}$;

$\mathbf{x} \in H_{\mathbf{x}_4}$, calculate $\mathbf{f}_{\mathbf{x}_4\mathbf{x}}$, add $\mathbf{f}_{\mathbf{x}_4\mathbf{x}} \Delta V_\mathbf{x} \Delta V_{\mathbf{x}_4}$ to $\mathbf{x}_4$, add $-\mathbf{f}_{\mathbf{x}_4\mathbf{x}} \Delta V_\mathbf{x} \Delta V_{\mathbf{x}_4}$ to $\mathbf{x}$.

Hence, in the process of calculating forces in horizons, the forces from dual-horizons are automatically done. Compared with the algorithm for OSB-PD given in [28], force expression: $\mathbf{f}_{\mathbf{x}\mathbf{x}'} - \mathbf{f}_{\mathbf{x}'\mathbf{x}}$ for point $\mathbf{x}$ with bond $\mathbf{x}\mathbf{x}'$ and force expression: $\mathbf{f}_{\mathbf{x}'\mathbf{x}} - \mathbf{f}_{\mathbf{x}\mathbf{x}'}$ for point $\mathbf{x}'$ with bond $\mathbf{x}'\mathbf{x}$ are calculated independently. Compared with the force summation in traditional PD, the bond force for every bond in DH-PD is calculated only once.

## 2.5. The bond force density $\mathbf{f}_{\mathbf{x}\mathbf{x}'}$

There mainly exist three types of peridynamics formulations [9,19], namely BB-PD, OSB-PD and NOSB-PD. Their key difference is the way they calculate the bond force density $\mathbf{f}_{\mathbf{x}'\mathbf{x}}$. The applications of DH-PD to the three types of peridynamics are direct [26]. The calibration of constitutive parameters with respect to the continuum model and some issues concerning the implementations will be discussed, for example, in BB-PD.

In BB-PD, the energy per unit volume in the body at a given time $t$ is given by [29]

$$W = \frac{1}{2} \int_{H_\mathbf{x}} w(\boldsymbol{\eta}, \boldsymbol{\xi}) \mathbf{d} V_{\boldsymbol{\xi}}, \tag{10}$$

where $w(\boldsymbol{\eta}, \boldsymbol{\xi})$ is the strain energy for bond $\boldsymbol{\xi}$. For the BB-PD theory, by enforcing the strain energy density being equal to the strain energy density in the classical theory of elasticity [29], we obtain the expression of the microelastic modulus $C(\delta)$ as

$$C(\delta) = \frac{3E}{\pi \delta^4 (1 - 2\nu)}, \tag{11}$$

where $\delta$ is the horizon radius associated to that material point, and $\nu$ is the Poisson's ratio. Note that the value of $C(\delta)$ takes half of the microelastic modulus $c$ used in the constant-horizon(CH) BB-PD since the bond energy for varying horizon is determined by both horizon and dual-horizon.

Let $w_0(\boldsymbol{\xi}) = C(\delta)s_0^2(\delta)\xi/2$ denote the work required to break a single bond, where $s_0(\delta)$ is the critical bond stretch. By breaking half of all the bonds connected to a given material point along the fracture surface and equalizing the breaking bonds energy with the critical energy release rate $G_0$ [11], we can get the expression between the critical energy release rate $G_0$ and the critical bond stretch $s_0(\delta)$ in three dimensions

$$s_0(\delta) = \sqrt{\frac{5G_0}{6E\delta}}. \tag{12}$$

Both the microelastic modulus $C(\delta)$ and the critical stretch $s_0(\delta)$ depend on the horizon radii for variable horizons.

In the implementation of the bond based peridynamics, fracture is introduced by removing points from the neighbor list once the bond stretch exceeds the critical bond stretch $s_0$. In order to specify whether a bond is broken or not, a history-dependent scalar valued function $\mu$ is introduced [29],

$$\mu(t, \boldsymbol{\xi}) = \begin{cases} 1 & \text{if } s(t', \boldsymbol{\xi}) < s_0 \text{ for all } 0 \leq t' \leq t, \\ 0 & \text{otherwise}. \end{cases} \tag{13}$$

The local damage at **x** is defined as

$$\phi(\mathbf{x}, t) = 1 - \frac{\int_{H_\mathbf{x}} \mu(\mathbf{x}, t, \boldsymbol{\xi}) dV_\xi}{\int_{H_\mathbf{x}} dV_\xi}. \tag{14}$$

For any material point with dual-horizon ($H'_\mathbf{x}$) and horizon ($H_\mathbf{x}$), the direct force $\mathbf{f}_{\mathbf{xx}'}$ and the reaction force $\mathbf{f}_{\mathbf{x}'\mathbf{x}}$ in Eq. (5) or (8) are computed by the following expressions, respectively

$$\mathbf{f}_{\mathbf{xx}'} = C(\delta_{\mathbf{x}'}) \cdot s_{\mathbf{xx}'} \cdot \frac{\boldsymbol{\eta} + \boldsymbol{\xi}}{\|\boldsymbol{\eta} + \boldsymbol{\xi}\|}, \quad \forall \mathbf{x}' \in H_\mathbf{x} \tag{15}$$

$$\mathbf{f}_{\mathbf{x}'\mathbf{x}} = C(\delta_\mathbf{x}) \cdot s_{\mathbf{xx}'} \cdot \frac{-(\boldsymbol{\eta} + \boldsymbol{\xi})}{\|\boldsymbol{\eta} + \boldsymbol{\xi}\|}, \quad \forall \mathbf{x}' \in H'_\mathbf{x}, \tag{16}$$

where $C(\delta_{\mathbf{x}'})$ and $C(\delta_\mathbf{x})$ are the microelastic modulus based on $\delta_{\mathbf{x}'}$ and $\delta_\mathbf{x}$ computed from Eq. (11) respectively, and $s_{\mathbf{xx}'}$ is the stretch between points **x** and **x**′.

For the other two types of peridynamics [19], comparing the equation of motion in traditional state-based PD with Eq. (9), the expression for bond force $\mathbf{f}_{\mathbf{xx}'}$ and $\mathbf{f}_{\mathbf{xx}'}$ can be obtained with ease. For more details please refer to [26].

## 3. The dual property of dual-horizon

Let $\mathcal{F}(\mathbf{x}, \mathbf{x}')$ be any expression depending on bond $\mathbf{xx}'$. The dual property of dual-horizon is that the double integration of the term $\mathcal{F}(\mathbf{x}, \mathbf{x}')$ in dual-horizon can be converted to the double integration of the term $\mathcal{F}(\mathbf{x}', \mathbf{x})$ in horizon, as shown in Eqs. (17) and (18). The key idea lies in that the term $\mathcal{F}(\mathbf{x}, \mathbf{x}')$ can be both interpreted in $H_\mathbf{x}$ and $H'_{\mathbf{x}'}$.

$$\sum_\Omega \left( \sum_{H'_\mathbf{x}} \mathcal{F}(\mathbf{x}, \mathbf{x}') \Delta V_{\mathbf{x}'} \right) \Delta V_\mathbf{x} = \sum_\Omega \left( \sum_{H_\mathbf{x}} \mathcal{F}(\mathbf{x}', \mathbf{x}) \Delta V_{\mathbf{x}'} \right) \Delta V_\mathbf{x} \qquad \text{Discrete form} \tag{17}$$

$$\int_\Omega \int_{H'_\mathbf{x}} \mathcal{F}(\mathbf{x}, \mathbf{x}') dV_{\mathbf{x}'} dV_\mathbf{x} = \int_\Omega \int_{H_\mathbf{x}} \mathcal{F}(\mathbf{x}', \mathbf{x}) dV_{\mathbf{x}'} dV_\mathbf{x} \qquad \text{Continuum form.} \tag{18}$$

**Proof.** Let $\Omega$ be discretized with $N$ voronoi tessellations (or other shape), as shown in Fig. 6. Each polygon is denoted with an index $i \in \{1, \ldots, N\}$, $\mathbf{x}_i$ is the coordinate for $i$'s center of gravity, $\Delta V_i$ is the volume associated to polygon $i$, $H_i$ and $H'_i$ are $i$'s horizon and dual-horizon, respectively. So

$$\Omega = \sum_{1 \leq i \leq N} \Delta V_i.$$

Let $\mathcal{F}(i, j) := \mathcal{F}(\mathbf{x}_i, \mathbf{x}_j)$ be any expression depending on bond $\mathbf{x}_i \mathbf{x}_j$. Consider the double summation of $\mathcal{F}(i, j)$ on $\Omega$.

$$\sum_{\Delta V_i \in \Omega} \left( \sum_{\Delta V_j \in H'_i} \mathcal{F}(i, j) \Delta V_j \right) \Delta V_i$$

$$= \sum_{1 \leq i \leq N} \left( \sum_{j \in H'_i} \mathcal{F}(i, j) \Delta V_j \right) \Delta V_i$$

$$= \sum_{j \in H'_1} \mathcal{F}(1, j) \Delta V_j \Delta V_1 + \sum_{j \in H'_2} \mathcal{F}(2, j) \Delta V_j \Delta V_2 + \cdots + \sum_{j \in H'_N} \mathcal{F}(N, j) \Delta V_j \Delta V_N. \tag{19}$$

In the third step, $j \in \{1, \ldots, N\}$ means $j$ belongs to that point's dual-horizon. Each term $\mathcal{F}(i, j) \Delta V_j \Delta V_i$ in $H'_i$ can be interpreted as term $\mathcal{F}(i, j) \Delta V_i \Delta V_j$ in $H_j$. Let us sum all terms in a way based on point $j$'s horizon $H_j$, where $j \in \{1, \ldots, N\}$

$$\sum_{j \in H'_1} \mathcal{F}(1, j) \Delta V_j \Delta V_1 + \sum_{j \in H'_2} \mathcal{F}(2, j) \Delta V_j \Delta V_2 + \cdots + \sum_{j \in H'_N} \mathcal{F}(N, j) \Delta V_j \Delta V_N$$

$$= \sum_{i \in H_1} \mathcal{F}(i, 1) \Delta V_1 \Delta V_i + \sum_{i \in H_2} \mathcal{F}(i, 2) \Delta V_2 \Delta V_i + \cdots + \sum_{i \in H_N} \mathcal{F}(i, N) \Delta V_N \Delta V_i. \tag{20}$$

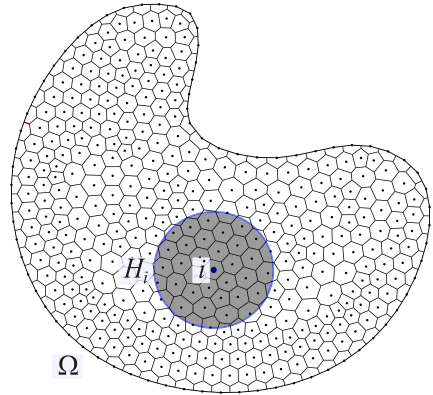

Fig. 6. The discretization of domain $\Omega$.

In the second step of Eq. (20), for example, $\sum_{i \in H_1} \mathcal{F}(i, 1) \Delta V_1 \Delta V_i$ means gathering all terms $j = 1$ in

$$\sum_{j \in H'_2} \mathcal{F}(2, j) \Delta V_j \Delta V_2 + \sum_{j \in H'_3} \mathcal{F}(3, j) \Delta V_j \Delta V_3 + \cdots + \sum_{j \in H'_N} \mathcal{F}(N, j) \Delta V_j \Delta V_N$$

$$\sum_{i \in H_1} \mathcal{F}(i, 1) \Delta V_1 \Delta V_i + \sum_{i \in H_2} \mathcal{F}(i, 2) \Delta V_2 \Delta V_i + \cdots + \sum_{i \in H_N} \mathcal{F}(i, N) \Delta V_N \Delta V_i$$

$$= \sum_{j \in H_1} \mathcal{F}(j, 1) \Delta V_j \Delta V_1 + \sum_{j \in H_2} \mathcal{F}(j, 2) \Delta V_j \Delta V_2 + \cdots + \sum_{j \in H_N} \mathcal{F}(j, N) \Delta V_j \Delta V_N$$

$$= \sum_{1 \leq i \leq N} \left( \sum_{j \in H_i} \mathcal{F}(j, i) \Delta V_j \right) \Delta V_i. \tag{21}$$

In the second step of Eq. (21), $i$ and $j$ are swapped, yielding

$$\sum_{1 \leq i \leq N} \left( \sum_{j \in H'_i} \mathcal{F}(i, j) \Delta V_j \right) \Delta V_i = \sum_{1 \leq i \leq N} \left( \sum_{j \in H_i} \mathcal{F}(j, i) \Delta V_j \right) \Delta V_i.$$

When $i$ and $j$ are replaced with $\mathbf{x}$ and $\mathbf{x}'$, respectively, we have

$$\sum_{\Omega} \left( \sum_{H'_{\mathbf{x}}} \mathcal{F}(\mathbf{x}, \mathbf{x}') \Delta V_{\mathbf{x}'} \right) \Delta V_{\mathbf{x}} = \sum_{\Omega} \left( \sum_{H_{\mathbf{x}}} \mathcal{F}(\mathbf{x}', \mathbf{x}) \Delta V_{\mathbf{x}'} \right) \Delta V_{\mathbf{x}}. \tag{22}$$

When $N \to \infty$ so that $\Delta V_i \to 0$, we have

$$\lim_{N \to \infty} \sum_{1 \leq i \leq N} \left( \sum_{j \in H'_i} \mathcal{F}(i, j) \Delta V_j \right) \Delta V_i = \int_{i \in \Omega} \int_{j \in H'_i} \mathcal{F}(i, j) \mathbf{d} V_j \mathbf{d} V_i.$$

Hence, the dual property of dual-horizon in the integral form is

$$\int_{\Omega} \int_{H'_{\mathbf{x}}} \mathcal{F}(\mathbf{x}, \mathbf{x}') \, \mathbf{d} V_{\mathbf{x}'} \mathbf{d} V_{\mathbf{x}} = \int_{\Omega} \int_{H_{\mathbf{x}}} \mathcal{F}(\mathbf{x}', \mathbf{x}) \, \mathbf{d} V_{\mathbf{x}'} \mathbf{d} V_{\mathbf{x}}. \tag{23}$$

Eq. (23) means the double integration of the term in dual-horizon can be converted to the double integration of the term with $\mathbf{x}$ and $\mathbf{x}'$ swapped in horizon.

When $\mathcal{F}(\mathbf{x}, \mathbf{x}') = \mathbf{f}_{\mathbf{xx}'}$, we have

$$\int_{\Omega} \int_{H_{\mathbf{x}}} \mathbf{f}_{\mathbf{xx}'} \, \mathbf{d} V_{\mathbf{x}'} \mathbf{d} V_{\mathbf{x}} = \int_{\Omega} \int_{H'_{\mathbf{x}}} \mathbf{f}_{\mathbf{x}'\mathbf{x}} \, \mathbf{d} V_{\mathbf{x}'} \mathbf{d} V_{\mathbf{x}}. \tag{24}$$

When $\mathcal{F}(\mathbf{x}, \mathbf{x}') = (\mathbf{x} + \mathbf{u}) \times \mathbf{f}_{\mathbf{xx}'}$, we have

$$\int_\Omega \int_{H_\mathbf{x}} (\mathbf{x} + \mathbf{u}) \times \mathbf{f}_{\mathbf{xx}'} \, dV_{\mathbf{x}'} dV_\mathbf{x} = \int_\Omega \int_{H'_\mathbf{x}} (\mathbf{x}' + \mathbf{u}') \times \mathbf{f}_{\mathbf{x}'\mathbf{x}} \, dV_{\mathbf{x}'} dV_\mathbf{x}. \tag{25}$$

### 3.1. Proof of basic physical principles

Two basic physical principles include the balance of the linear momentum and the balance of the angular momentum. We showed in [26] that DH-PD fulfills these basic principles. In this manuscript, we provide a shorter proof based on the dual property of the dual-horizon.

**Balance of linear momentum**

The internal forces shall satisfy the balance of linear momentum for any bounded body $\Omega$ given in form by

$$\int_\Omega (\rho \ddot{\mathbf{u}}(\mathbf{x}, t) - \mathbf{b}(\mathbf{x}, t)) dV_\mathbf{x}$$
$$= \int_\Omega \int_{H'_\mathbf{x}} \mathbf{f}_{\mathbf{xx}'}(\boldsymbol{\eta}, \boldsymbol{\xi}) \, dV_{\mathbf{x}'} dV_\mathbf{x} - \int_\Omega \int_{H_\mathbf{x}} \mathbf{f}_{\mathbf{x}'\mathbf{x}}(-\boldsymbol{\eta}, -\boldsymbol{\xi}) \, dV_{\mathbf{x}'} dV_\mathbf{x}$$
$$= \mathbf{0}. \tag{26}$$

**Proof.** Based on Eq. (24), Eq. (26) is apparently satisfied, so the linear momentum is conserved. □

**Balance of angular momentum**

To satisfy the balance of angular momentum for any bounded body $\Omega$, it is required that

$$\int_\Omega \mathbf{y} \times (\rho \ddot{\mathbf{u}}(\mathbf{x}, t) - \mathbf{b}(\mathbf{x}, t)) dV_\mathbf{x}$$
$$= \int_\Omega \mathbf{y} \times \left( \int_{H'_\mathbf{x}} \mathbf{f}_{\mathbf{xx}'}(\boldsymbol{\eta}, \boldsymbol{\xi}) \, dV_{\mathbf{x}'} - \int_{H_\mathbf{x}} \mathbf{f}_{\mathbf{x}'\mathbf{x}}(-\boldsymbol{\eta}, -\boldsymbol{\xi}) \, dV_{\mathbf{x}'} \right) dV_\mathbf{x}$$
$$= \int_\Omega \int_{H'_\mathbf{x}} \mathbf{y} \times \mathbf{f}_{\mathbf{xx}'}(\boldsymbol{\eta}, \boldsymbol{\xi}) \, dV_{\mathbf{x}'} dV_\mathbf{x} - \int_\Omega \int_{H_\mathbf{x}} \mathbf{y} \times \mathbf{f}_{\mathbf{x}'\mathbf{x}}(-\boldsymbol{\eta}, -\boldsymbol{\xi}) \, dV_{\mathbf{x}'} dV_\mathbf{x}$$
$$= \mathbf{0}. \tag{27}$$

For simplicities, let $\mathbf{f}_{\mathbf{xx}'}$ represent $\mathbf{f}_{\mathbf{xx}'}(\boldsymbol{\eta}, \boldsymbol{\xi})$, and $\mathbf{f}_{\mathbf{x}'\mathbf{x}}$ represent $\mathbf{f}_{\mathbf{x}'\mathbf{x}}(-\boldsymbol{\eta}, -\boldsymbol{\xi})$.

**Proposition.** *In the dual-horizon peridynamics, suppose a constitutive model of the form*

$$\mathbf{f} = \hat{\mathbf{f}}(\mathbf{y}, \Lambda) \tag{28}$$

*where* $\hat{\mathbf{f}} : \mathcal{V} \to \mathcal{V}$ *is bounded and Riemann-integrable on Hilbert function space and $\mathcal{V}$ is the vector state; $\Lambda$ denotes all variables other than the current deformation vector state.*
*If*

$$\int_{H_\mathbf{x}} \mathbf{y}\langle \mathbf{x}' - \mathbf{x} \rangle \times \mathbf{f}_{\mathbf{x}'\mathbf{x}} \, dV_{\mathbf{x}'} = 0 \qquad \forall \mathbf{y} \in \mathcal{V}, \tag{29}$$

*then the balance of angular momentum, Eq. (27) holds for any deformation of $\Omega$ for any given constitutive model.*

**Proof.** Based on Eq. (25), Eq. (27) can be written as

$$\int_\Omega \mathbf{y} \times (\rho \ddot{\mathbf{u}}(\mathbf{x}, t) - \mathbf{b}(\mathbf{x}, t)) dV_\mathbf{x}$$
$$= \int_\Omega \mathbf{y} \times \left( \int_{H'_\mathbf{x}} \mathbf{f}_{\mathbf{xx}'}(\boldsymbol{\eta}, \boldsymbol{\xi}) \, dV_{\mathbf{x}'} - \int_{H_\mathbf{x}} \mathbf{f}_{\mathbf{x}'\mathbf{x}}(-\boldsymbol{\eta}, -\boldsymbol{\xi}) \, dV_{\mathbf{x}'} \right) dV_\mathbf{x}$$

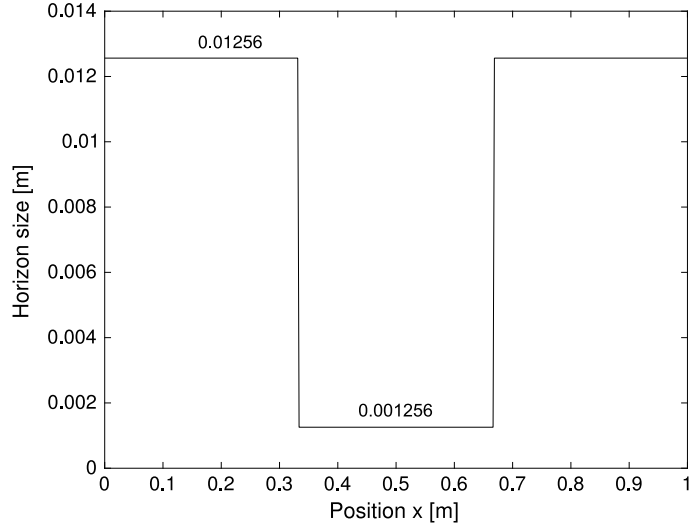

Fig. 7. The horizon sizes distribution.

$$\begin{aligned}
&= \int_\Omega \int_{H'_x} \mathbf{y} \times \mathbf{f}_{xx'}(\eta, \xi) \, dV_{x'} dV_x - \int_\Omega \int_{H_x} \mathbf{y} \times \mathbf{f}_{x'x}(-\eta, -\xi) \, dV_{x'} dV_x \\
&= \int_\Omega \int_{H_x} \mathbf{y}' \times \mathbf{f}_{x'x}(-\eta, -\xi) \, dV_{x'} dV_x - \int_\Omega \int_{H_x} \mathbf{y} \times \mathbf{f}_{x'x}(-\eta, -\xi) \, dV_{x'} dV_x \\
&= \int_\Omega \int_{H_x} (\mathbf{y}' - \mathbf{y}) \times \mathbf{f}_{x'x}(-\eta, -\xi) \, dV_{x'} dV_x \\
&= \mathbf{0}.
\end{aligned} \qquad (30)$$

The expression in step 4 is equal to zero in all three types of peridynamics (see proof in [19]), so the angular momentum is conserved. □

It can be seen that the conservation of angular momentum somehow depends **only** on the horizon; the dual-horizon is not involved. The latter is only needed in Eqs. (5) and (8). This conclusion can be also illustrated for the BB-PD and OS-PD since the internal forces $\mathbf{f}_{xx'}$ and $\mathbf{f}_{x'x}$ are parallel to the bond vector in the current configuration, see Fig. 2.

## 4. Wave propagation in 1D homogeneous bar

Consider a 1D bar with length of $L = 1$ m, one third of the bar in the middle is discretized with $\Delta x_1 = 4.19 \times 10^{-3}$ m and the other parts discretized with $\Delta x_2 = 10 \Delta x_1$. The horizon size is selected as $\delta_i = 3.015 \Delta x_i$, as shown in Fig. 7. Both ends of the bar are free. The initial displacement field is given by

$$u(x, 0) = 0.002 \exp\left(-\frac{(x - 0.5L)^2}{1000}\right). \qquad (31)$$

The bar is solved with DH-PD (Case I) and traditional PD formulation (Case II). The wave profiles at different times are shown in Fig. 8 for DH-PD and Fig. 9 for traditional PD. It can be seen the horizon size interfaces have limited influence on the wave profiles in Case I, while the spurious wave emerged for traditional PD formulation. Two points $x = 0.32$ m and $x = 0.34$ m are selected to analyze the wave profile, and the displacements at different time are shown in Fig. 10. It shows that the incident wave passed the interface and the magnitude of reflected wave is smaller than 3.17% of the magnitude of incident wave.

On the other hand, for the traditional PD without any treatment on variable horizons, when the incident wave approached the interface, the deformation of the fine material points caused the ghost forces among the coarse material points. The ghost forces give rise to the spurious wave in the domain with coarse material points, as shown in Fig. 9 ($T = 0.2$ s). In addition, the wave in the fine material points is hard to pass the interface but to be reflected due to the fact that its horizon size is much smaller than that of coarse material points.

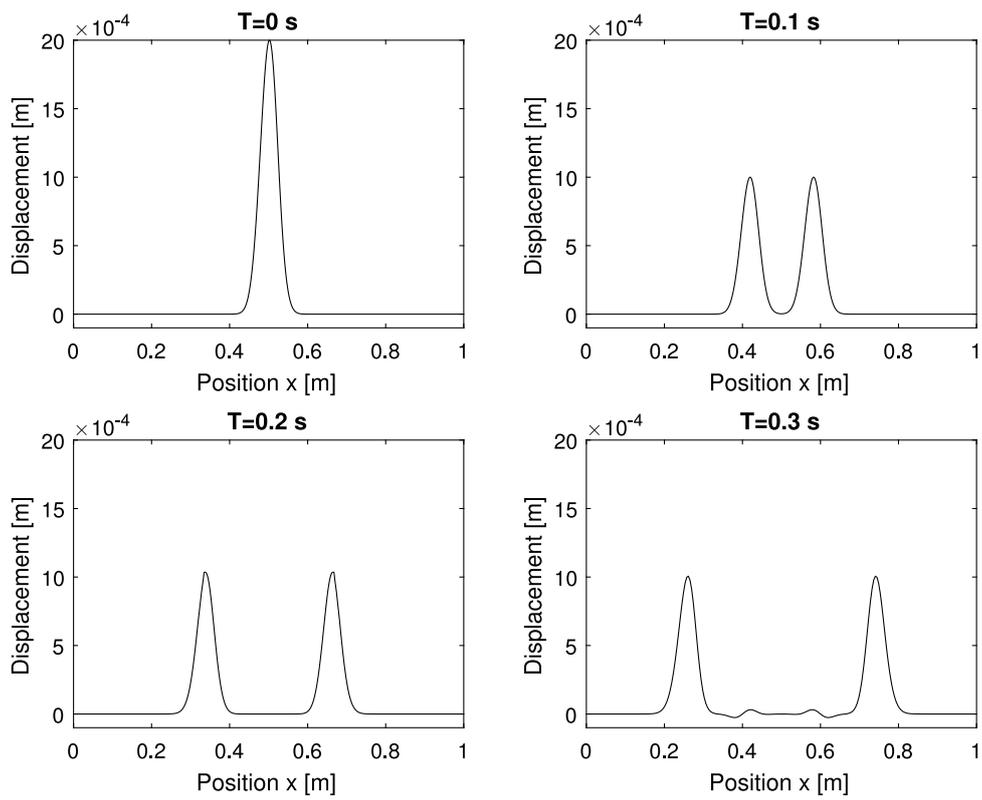

Fig. 8. The wave profiles for Case I at different times.

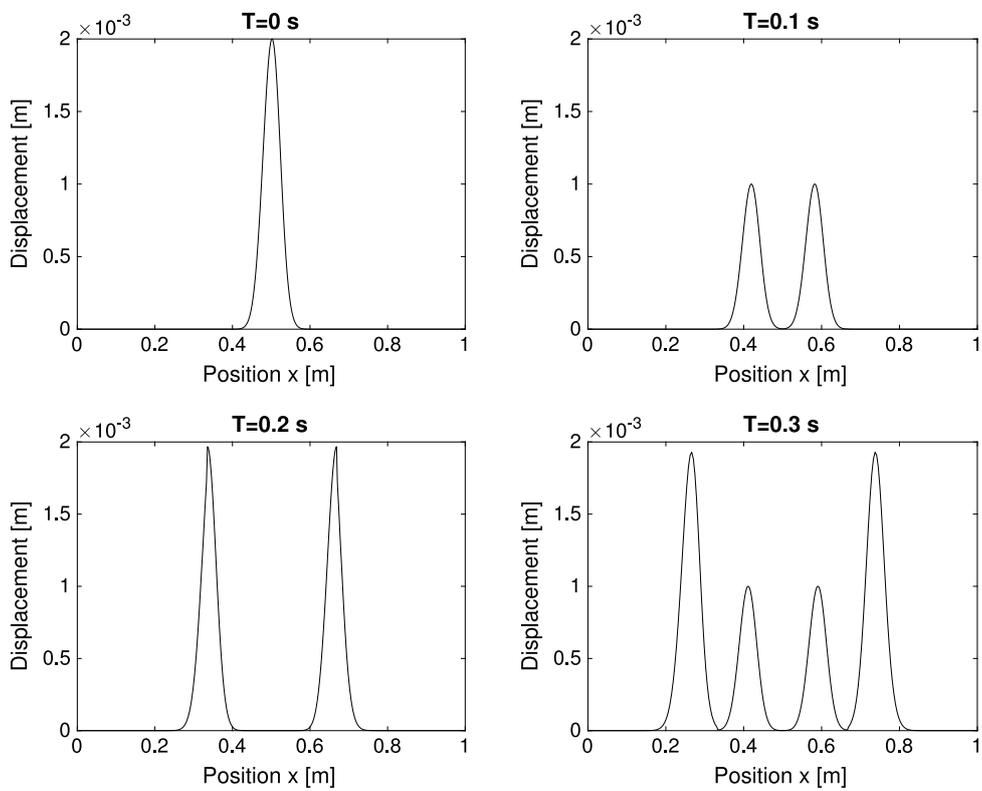

Fig. 9. The wave profiles for Case II at different times.

## 5. Multiple materials

The material interface problem in peridynamics was formally discussed in [30], where the authors analyzed the convergence of peridynamics for heterogeneous media, and found that the operator of linear peridynamics diverges in the presence of material interfaces. Some approaches have been proposed to deal with this issue. In Ref. [19], the interfaces and mixtures between multiple materials are accounted for by using separate influence functions for each materials. For example, if $\mathbf{x}$ is near an interface between two materials denoted by (1) and (2), two separate influence functions $\omega^{(1)}$ and $\omega^{(2)}$ can be defined such that each vanishes at points $\mathbf{x} + \boldsymbol{\xi}$ outside its associated material. The force state is given by

$$\mathbf{T}_{\mathbf{xx}'} = \omega^{(1)} \langle \boldsymbol{\xi} \rangle P_{\mathbf{x}}^{(1)} \mathbf{K}^{-1} \boldsymbol{\xi} + \omega^{(2)} \langle \boldsymbol{\xi} \rangle P_{\mathbf{x}}^{(2)} \mathbf{K}^{-1} \boldsymbol{\xi} \tag{32}$$

where $P_{\mathbf{x}}^{(1)}$ and $P_{\mathbf{x}}^{(1)}$ are the first Piola–Kirchhoff stress from the respective constitution models and respective deformation gradient tensors $\mathbf{F}^{(1)}$ and $\mathbf{F}^{(2)}$, $\mathbf{K}$ is the shape tensor. In Ref. [30], a peridynamic interface model was proposed. However, such issue can be handled in a simpler way in DH-PD. In the DH-PD formulation, any point $\mathbf{x}$ has a force state $\mathbf{T}_{\mathbf{x}\square}$, which acts like a force potential field applied to its nearby neighbors in $H_{\mathbf{x}}$. We assume $\mathbf{x}$ imposes reaction bond forces to its neighbors no matter what the neighbor's material type is. The bond force can be determined by one side or both sides. For examples, in the BB-PD, microelastic modulus $C(\delta)$ and critical material stretch $s(\delta)$ are the primary parameters. For two points $i$ and $j$ with initial coefficient $[C(\delta_i), s(\delta_i)]$ (abbreviated as $[C_i, s_i]$) and $[C(\delta_j), s(\delta_j)]$, respectively, when taking into consideration of the combination of the initial coefficients, the alternate forms can be formulated as the arithmetic average or geometric average of microelastic moduli or critical stretches.

For the OSB-PD and NOSB-PD in multiple materials, the force state $t_{\mathbf{xx}'}$ and $T_{\mathbf{xx}'}$ can be calculated similarly to the single homogeneous material and hence, it is not necessary to distinguish between homogeneous and heterogeneous materials.

### 5.1. 1D bar tensile test

The length of the bar is 1 m, discretized with grid spacing $\Delta x = 0.01, 0.005$ m. The elastic modulus and density are $E_1 = 1, E_2 = 2$ and $\rho = 1$, respectively. Boundary conditions are applied at 3 material points at each end, i.e. fixed boundary condition at left end (blue points in Fig. 11) and displacement boundary conditions of $\Delta s = 1e-4$ m (red points in Fig. 11). The damping coefficient is $\alpha = 5e-3$ m. Two cases were tested. Case I is based on the traditional constant-horizon BB-PD but without any treatment for multiple materials, while Case II on dual-horizon BB-PD with variable horizons. The equilibrium state is obtained at $t = 0.8$ s. The bar internal sectional force is approximated by summing the bond force at one side given by

$$f_i = \delta_i \sum_{j \in H_i, j > i} C_i s_{ij} \Delta V_j \tag{33}$$

where $s_{ij}$ is the bond stretch for bond $ij$, $\delta_i$ is the radius of horizon $H_i$. Due to the error of discretization, Eq. (33) cannot sum the sectional force from material point $i$ accurately; when right half of $H_i$ contains just 3 neighbors and volume correction is considered, there are 2.5 material points in Eq. (33), which corresponds to 5/6 of the theoretical sectional force. For Case I, the force discontinuity in the interface of different point spacing is observed in Fig. 12. The force solution diverts from the analytical solution. The reason for the result is that the microelastic modulus is determined by one point. Consider a point $i$ close to the interface of two materials with microelastic modulus $C_i$, the equilibrium of $i$ requires the total force from left side being equal to that from right side. As a result, the displacement field in left half and right half horizon are anti-symmetrical with respect to $i$. Hence, there is no discontinuity in the strain field at the interface. At last, the spurious equilibrium displacement and force field are obtained.

However, for Case II modeled with DH-PD, the force solution and displacement solution agree well with the analytical solutions, respectively, see Fig. 13. Both the material interface and point spacing interface have minimal influence on the final solution, though small variation is observed at the interfaces of material or grid spacing. The variation of the sectional force arises from the nonlocality in peridynamics. In PD, the force can transmit over finite distance, the equilibrium of a point means the forces from horizon and dual-horizon are canceled. In this sense, the equilibrium is determined by the domains of horizon and dual-horizon, two nearby material points may have different domains, therefore, the jumping of internal sectional force is reasonable. This is essentially different from the ghost force which is unbalanced.

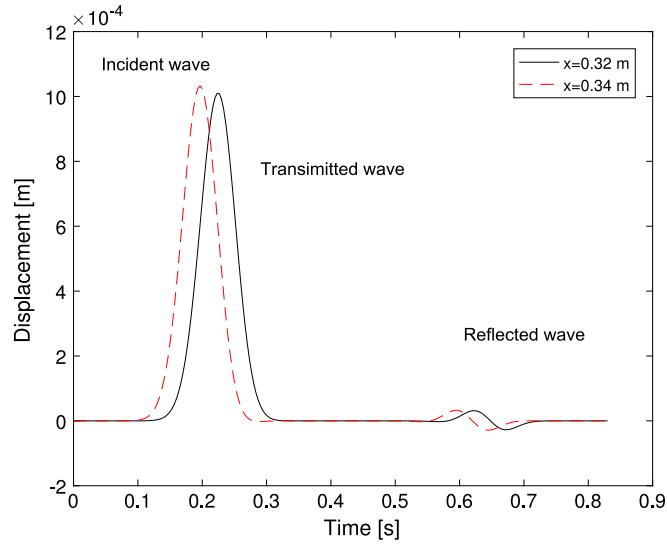

Fig. 10. The wave for points $x = 0.32$ m and $x = 0.34$ m.

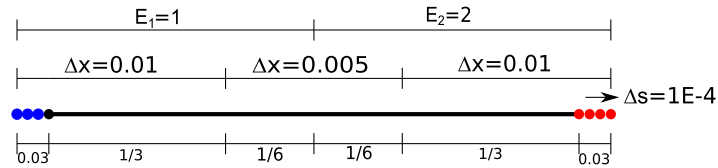

Fig. 11. The setup of 1D bar.

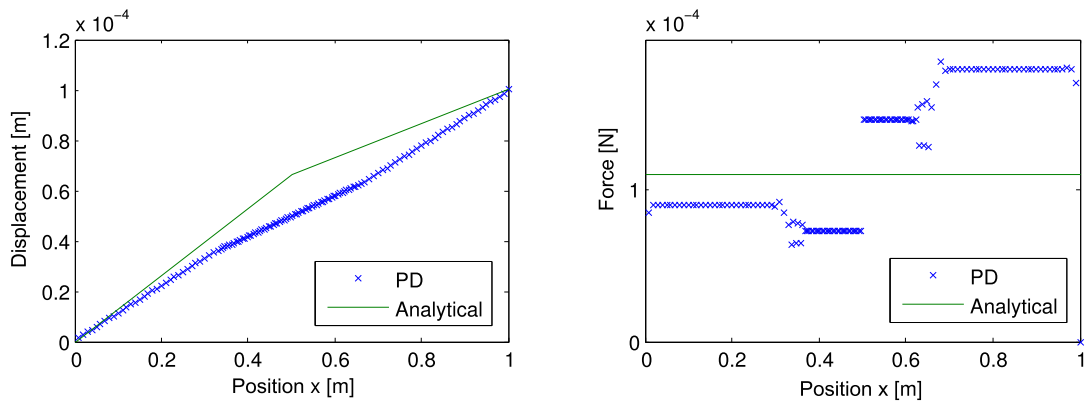

Fig. 12. The displacement and sectional force for Case 1.

In conclusion, we show for a simple 1D example that DH-PD has the potential to effectively deal with multiple materials.

### 5.2. Crack propagation in heterogeneous materials

Consider a plate subjected to constant velocity traction at both the top and down edges, as shown in Fig. 14. The matrix is filled with many randomly distributed inclusions; all inclusions have the same material parameters. The material parameters for matrix include $E_1 = 72$ GPa, $v_1 = 1/3$, $G_0 = 40$ J/m$^2$. The parameters for the inclusions are $E_2 = 144$ GPa, $v_2 = 1/3$, $G_0 = 80$ J/m$^2$. The plate is discretized into 40,000 material points with point spacing of $\Delta x = 1.5e{-}5$ m. The horizon radius is $\delta = 3.015\Delta x$. The material's critical bond stretch is calculated based on Eq. (12), i.e. $s_1 = s_2 \approx 4.64e{-}3$. Three cases are conducted to show the capability of DH-PD in dealing with composite materials. The velocity of traction on boundary is $v_{bc} = 0.5$ m/s for Case I, II and $v_{bc} = 2$ m/s for Case III. Two different heterogeneous structures are tested; the heterogeneous structures for Case II and III coincide.

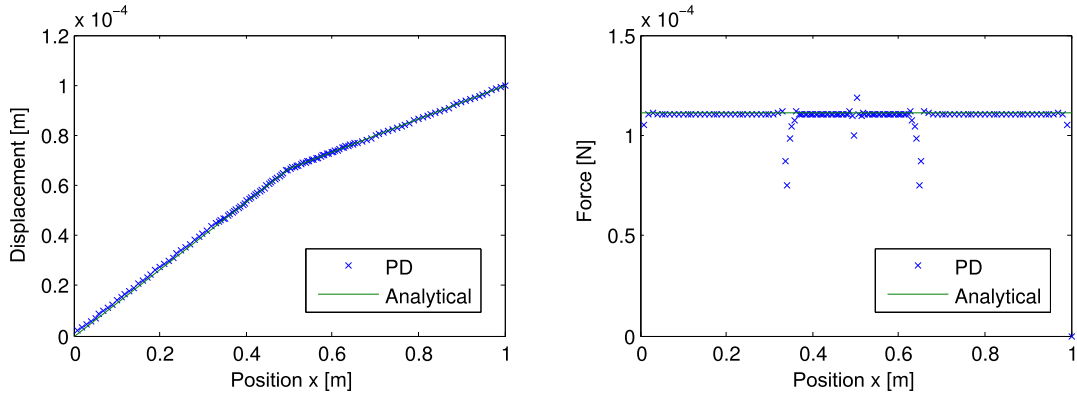

Fig. 13. The displacement and sectional force for Case 2.

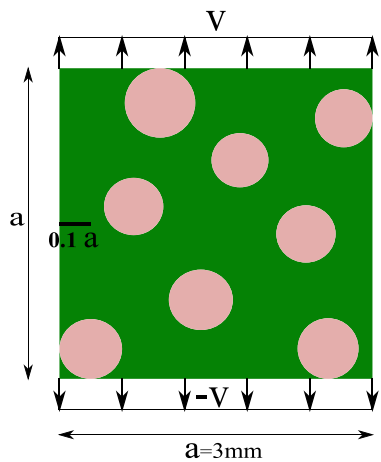

Fig. 14. The setup of the plate with inclusions.

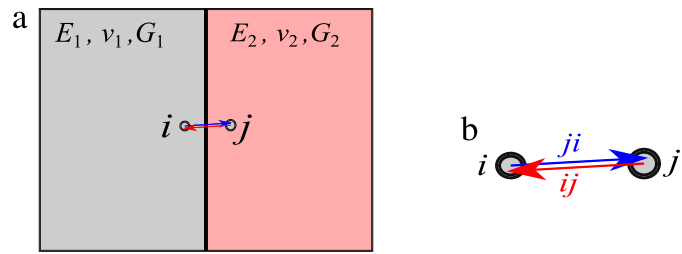

Fig. 15. The bonds in composite materials. Bond $ji$ depends on $i$. When calculating $i$, all neighbors in $H_i$ are pretended to have the same parameters as $i$. Bond $ij$ depends on $j$, when calculating $j$, all neighbors in $H_j$ are pretended to have the same parameters as $j$. Bond $ij$ and $ji$ are broken separately.

We use a simple damage rule to take into account the crack propagation in heterogeneous materials. In the framework of DH-PD, the bond and dual-bond are broken separately. For example, as shown in Fig. 15, bond $ji$ is interpreted as $i$ exerting force on $j$, and the strength of bond $ji$ is assumed to depend only on $i$ parameters, e.g. $i$'s critical stretch; the same applies to bond $ij$. Figs. 16(b) and 16(c) show when initial crack tip is in the inclusion (Fig. 16(c)), the crack will firstly propagate in the inclusions; when initial crack tip is in the matrix, the crack firstly propagates in the matrix. We also note two phenomena when the load rate is increased:

(1) The number of crack branches is increased. Such a phenomenon is well known in dynamic fracture.
(2) The crack propagates through the inclusions. This tendency has been also observed experimentally, i.e. the number of aggregates/particle cracks in a weaker matrix material is increased with increasing strain rate/load rate.

14　　　　　　　　　　　　　　　H. Ren et al. / Comput. Methods Appl. Mech. Engrg. 318 (2017) 0–20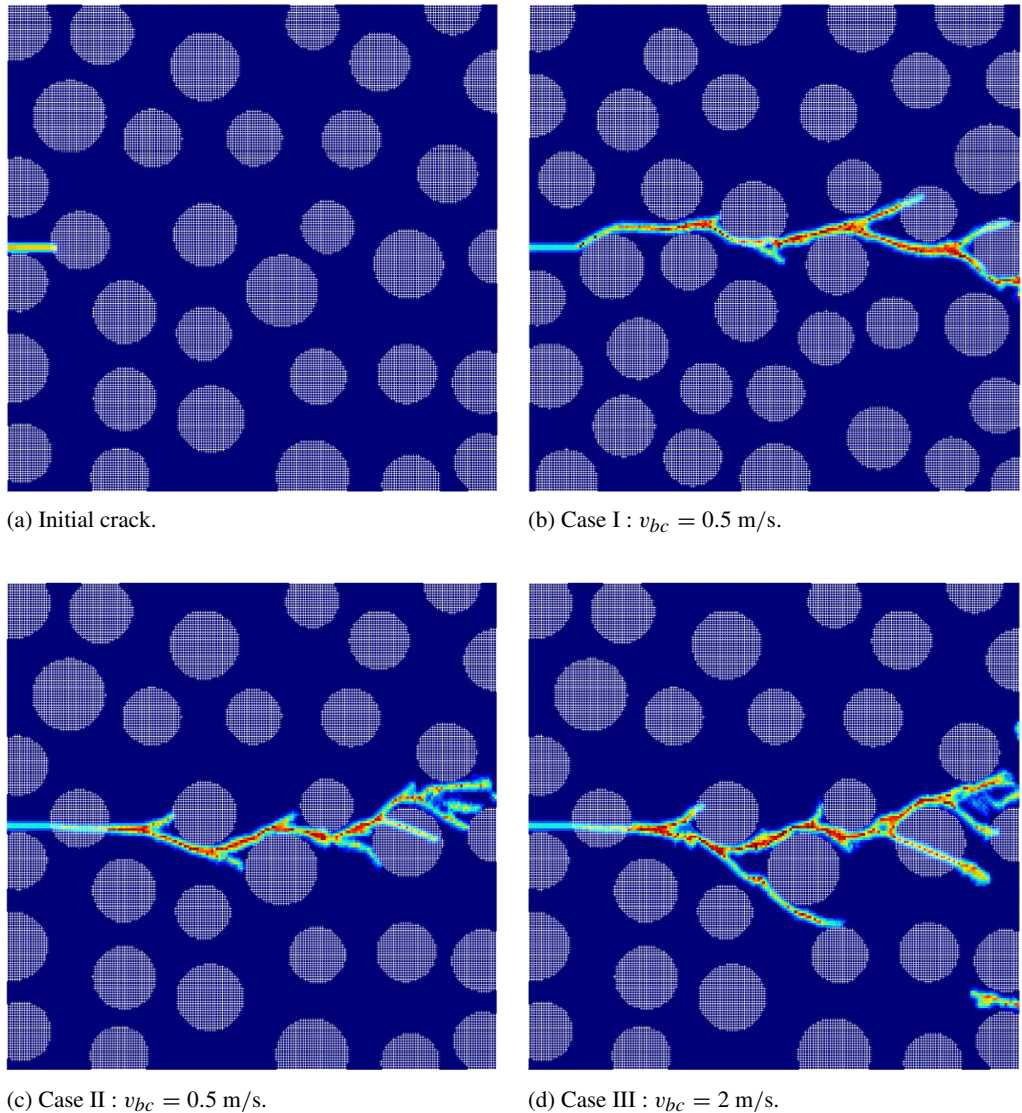

(a) Initial crack.　　　　　　　　　　　　　　(b) Case I : $v_{bc} = 0.5$ m/s.

(c) Case II : $v_{bc} = 0.5$ m/s.　　　　　　　　(d) Case III : $v_{bc} = 2$ m/s.

Fig. 16. The crack patterns in multiple-material plate.

## 6. Simulation of the Kalthoff–Winkler experiment

The Kalthoff–Winkler experiment is a classical benchmark problem studied by [31–37]. The setup is depicted in Fig. 19; the thickness of the specimen is 0.01 m. In the experiment, the evolution of crack pattern was observed to be dependent on the impact loading velocity. For a plate made of steel 18Ni1900 subjected to an impact loading at the speed of $v_0 = 32$ m/s, brittle fracture was observed [38]. The crack propagates from the end of the initial crack at an angle around 70° with respect to the initial crack direction. In [38], the BB-PD with constant horizon was used to test this example. For convenience of comparison, the present dual-horizon formulation is also applied to the bond based peridynamics to test the example. The material parameters used are the same as in [38], i.e. the elastic modulus $E = 190$ GPa, $\rho = 7800$ kg/m$^3$, $\nu = 0.25$ and the energy release rate $G_0 = 6.9 \times 10^4$ J/m$^2$. The impact loading was imposed by applying an initial velocity at $v_0 = 22$ m/s to the first three layers of material points in the domain as shown in Fig. 19; all other boundaries are free. For brevity, the surface correction method [39] or fictitious boundary [40] are not employed.

In order to create the irregular material points distribution in the computational domain, we utilize the Abaqus software to mesh the geometrical model with different mesh seeds setting, and then convert the finite element mesh into material points. For example, in the case of triangle element or tetrahedron element, each element's area(or volume) is evenly allocated to its nodes, as shown in Figs. 17 and 18. Every node (acted as material point) is associated with a unique volume or mass. The material point size $\Delta x$ is the equivalent diameter, which is calculated based on its volume by assuming the material point is circle or sphere. Six cases as shown in Table 1, are tested. Case I, II are in two dimensions with the same material points distribution where the ratio $\Delta x_{max}/\Delta x_{min} = 3.1$, while Case III–VI in three dimensions with particle size ratio $\Delta x_{max}/\Delta x_{min} = 4.4$. Case I and III are based on the traditional constant-horizon peridynamics (CH-PD), while Case II and IV are carried out with the DH-PD. Case V is conducted to test the influence of spurious wave reflections when variable horizons is used in traditional PD. Case VI is used to test the effect of volume correction given in Appendix A. Two different discretizations (for I, II and III–VI, respectively) with a bias towards a 'wrong' crack path. In Case II and IV–VI, the radius of each material point's horizon is selected as three times of the material point size, e.g. $\delta_i = 3.015 \Delta x_i$, while in Case I and III $\delta = 3.015 \Delta x_{max}$. The bond microelastic modulus $C(\delta_\mathbf{x})$ is calculated based on Eq. (11).

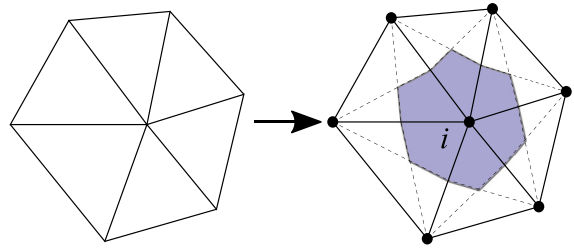

Fig. 17. Triangle elements to material points.

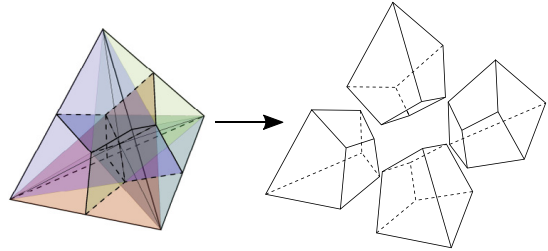

Fig. 18. Tetrahedron element to material points.

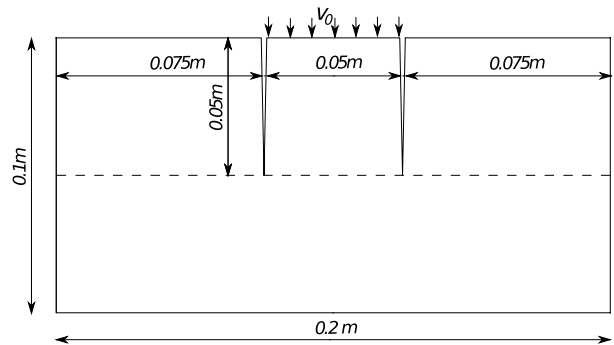

Fig. 19. Kalthoff–Winkler's experimental setup.

The angle of the crack propagation with respect to the original crack direction is shown in Table 2. For relatively low speed impact (e.g. ≤30 m/s), the crack angle in [41,42] is approximately 70°. Table 2 shows that result by DH-PD is better than that by CH-PD. The results in Fig. 20 and Table 2 show that the crack pattern both in 2D and 3D cases by the traditional constant-horizon bond based peridynamics differs from the experiment, while the crack pattern obtained by the DH-PD agrees well with the experiment besides the introduction of a mesh bias towards an 'incorrect' crack path. The comparison of Case IV and Case V shows the crack pattern is affected by the spurious wave reflections. The comparison of Case IV and Case VI shows the volume correction has some positive effect on the crack propagation, therefore, the volume correction is recommended.

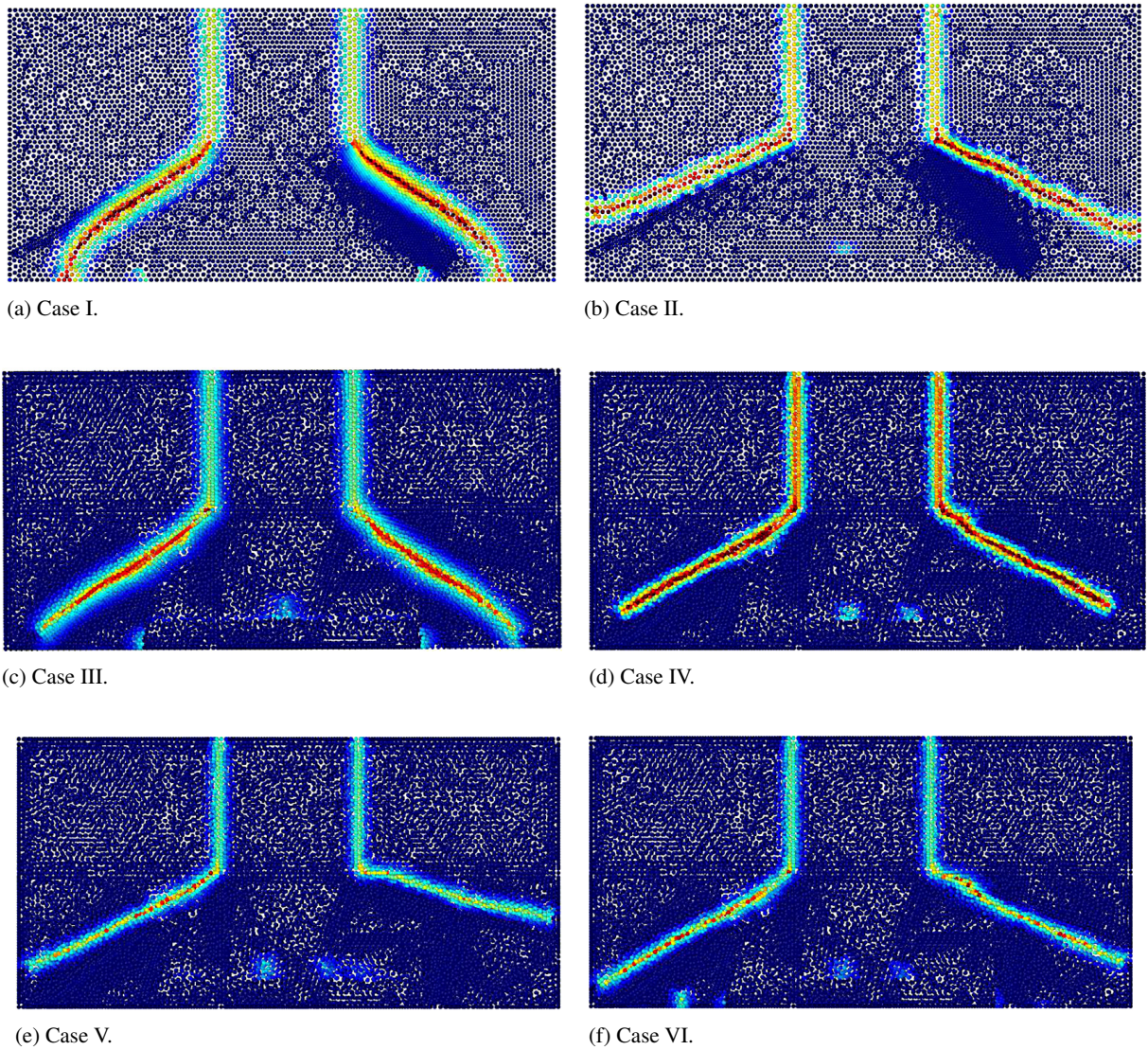

Fig. 20. The crack patterns in the simulations of Kalthoff–Winkler experiment.

Table 1
The parameters in 4 simulations of Kalthoff–Winkler experiment.

| Case | $\delta_{max}/\delta_{min}$ | Type | Volume correction |
|---|---|---|---|
| I | 1 | CH-PD | Yes |
| II | 3.1 | DH-PD | Yes |
| III | 1 | CH-PD | Yes |
| IV | 4.4 | DH-PD | Yes |
| V | 4.4 | CH-PD | Yes |
| VI | 4.4 | DH-PD | No |

Table 2
The angle of crack propagation with respect to the original crack.

| | Case I | Case II | Case III | Case IV | Case V | Case VI |
|---|---|---|---|---|---|---|
| Left | 56.1° | 67.6° | 56.4° | 61.4° | 62.5° | 58.2° |
| Right | 57.6° | 64.7° | 56.3° | 62.7° | 76.1° | 65.3° |

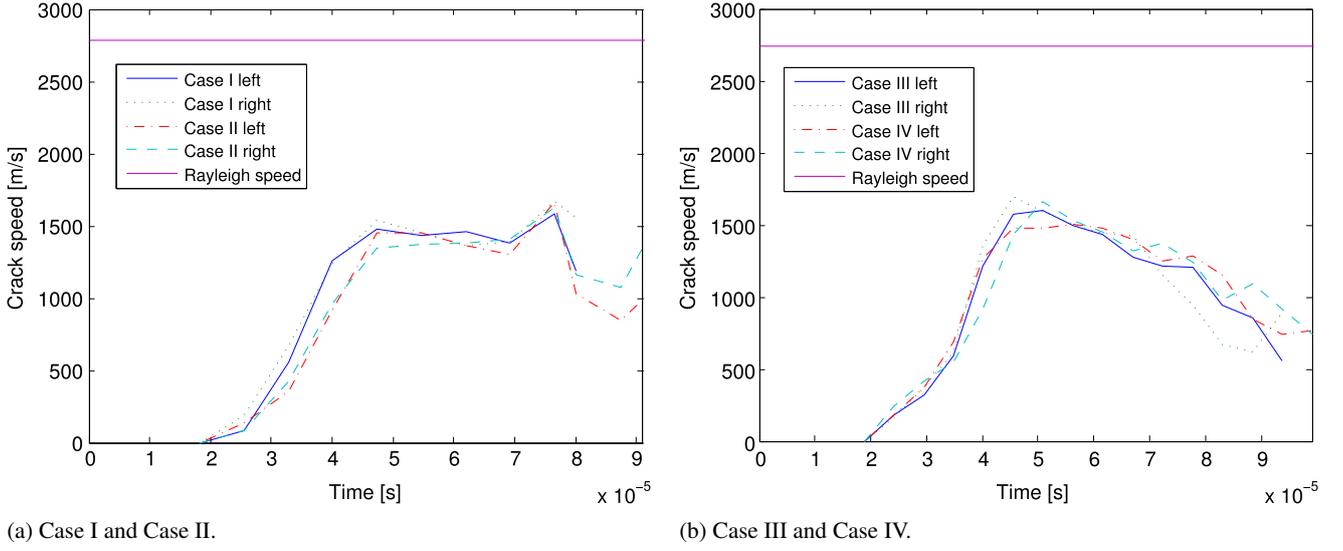

(a) Case I and Case II.    (b) Case III and Case IV.

Fig. 21. The crack propagation speed in the simulations of Kalthoff–Winkler experiment.

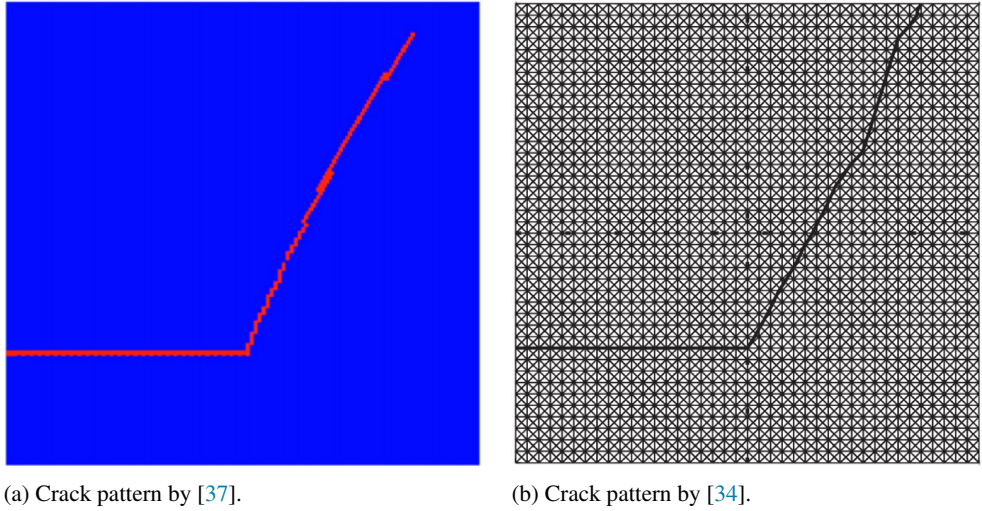

(a) Crack pattern by [37].    (b) Crack pattern by [34].

Fig. 22. The crack pattern of Kalthoff–Winkler experiment by other methods.

The crack propagation speed is computed by

$$V_{l-0.5} = \frac{\|\mathbf{x}_l - \mathbf{x}_{l-1}\|}{t_l - t_{l-1}}, \qquad (34)$$

where $\mathbf{x}_l$ and $\mathbf{x}_{l-1}$ are the positions of the crack tip at the times $t_l$ and $t_{l-1}$, respectively. An expression for the Rayleigh wave speed $c_R$ [43] is given as

$$\frac{c_R}{c_s} = \frac{0.87 + 1.12\nu}{1 + \nu}, \qquad (35)$$

where $\nu$ is the Poisson's ratio, $c_s = \sqrt{\mu/\rho}$ is the shear wave speed and $\mu$ is the shear modulus. The crack starts to propagate at 20 µs. The crack speed in all four cases simulation is shown in Fig. 21. The maximum crack propagation speeds are 1669 m/s, 1658 m/s, 1699 m/s, 1664 m/s for Case I, II, III, IV, respectively. All of them are within the limit of 75% of Rayleigh speed. The crack angle of Kalthoff–Winkler experiment by other methods [37,34] are 65.1° and 63.8° as shown in Figs. 22(a) and Fig. 22(b), respectively. Therefore, it can be concluded that with the present formulation, the random material points distribution has limited influence on the crack propagation when simulated by DH-PD.

## 7. Conclusions

This paper contributes to the development of dual-horizon peridynamics. The dual property of dual-horizon, a universal principle for dual-horizon, was proved. Based on the dual property of dual-horizon, we re-proved that the DH-PD fulfills both the balance of momentum and balance of angular momentum. Two issues of crack pattern on random material points distribution and multiple materials are discussed.

Several numerical examples were presented to demonstrate the capabilities of DH-PD in dealing with issue of ghost forces, multiple materials problems, and crack stability problems on random particle arrangement. The first numerical example shows that spurious wave reflection happened in the traditional bond based peridynamics, while the DH-PD can handle it with ease. The second numerical example shows that DH-PD can handle the multiple materials problem with minimal modification on traditional PD. The last numerical example shows that the crack patterns simulated by bond based DH-PD are stable and agrees well with the experiment result despite the irregular material points distribution, while that by traditional bond based PD are affected by the irregular material points distribution.

There are several advantages of DH-PD over constant horizon PD.

(1) The number of neighbors for each material point in DH-PD is less than that in constant-horizon PD when using an inhomogeneous discretization of the computational domain, therefore reduced the computational cost greatly. For certain small material points, it has very limited neighbors but may has many dual-neighbors as it is contained in nearby point' horizons; in the constant-horizon peridynamics, the small material point may have significantly large neighbor list, as discussed in Section 2.1.

(2) The DH-PD improves the computational efficiency in force summations since the force density in DH-PD is only calculated once while the force density in traditional PD is calculated twice, as shown in Section 2.4.

(3) The DH-PD can deal with the issue of multiple materials with minimal modification of the traditional peridynamics.

(4) The DH-PD can improve the stability of crack propagation and reduce the width of the smeared fractures compared with the constant-horizon peridynamics when using a inhomogeneous discretization of the computational domain.

## Acknowledgments

The authors acknowledge the supports from the National Basic Research Program of China (973 Program: 2011CB013800) and NSFC (51474157), Shanghai Municipal Commission of Science and Technology (16QA1404000), the Ministry of Science and Technology of China (SLDRCE14-B-31) and the ERC-CoG (Computational Modeling and Design of Lithium-ion Batteries (COMBAT)).

## Appendix

*A.1. Volume correction*

In order to account for the effect of material points partly falling inside the horizon, some volume correction techniques are utilized to improve the accuracy of force summation in horizon. The commonly used volume correction method in constant-horizon peridynamics is [28]

$$v_{\mathbf{x}_p - \mathbf{x}_i} = \begin{cases} -\dfrac{\|\mathbf{x}_p - \mathbf{x}_i\|}{2r_i} + \left(\dfrac{\delta}{2r_i} + \dfrac{1}{2}\right) & \text{if } \delta - r_i \leq \|\mathbf{x}_p - \mathbf{x}_i\| \leq \delta, \\ 1 & \text{if } \|\mathbf{x}_p - \mathbf{x}_i\| \leq \delta - r_i, \\ 0 & \text{otherwise,} \end{cases} \quad (A.1)$$

where $r_i = \Delta/2$, $\Delta$ is the spacing between material points, or the material point size. In the dual-horizon peridynamics, as any material point has its unique horizon, the volume ratio of big material point with small material point can be big. For example, one case as shown in Fig. 23, the center of material point $j$ is falling outside of $i$'s horizon but with certain nontrivial part (green domain in Fig. 23) of $j$ belonging to $H_i$. In order to improve the accuracy of integration in $H_i$, the green domain in Fig. 23 should be added to calculate the reaction force from $j$. As the material point is generated from finite element, the shape of each material point is irregular, the material point size can be estimated by its volume. Let $r_i$ denote material point $i$'s radius in the shape of circle or sphere. Let two circles (or spheres) of radii $R$ and $r$ be located along the $x$-axis centered at $(0, 0, 0)$ and $(d, 0, 0)$, respectively, as shown in Fig. 24. The intersection area $A$ (or volume $V$) between two circles (or spheres) is

$$A = r^2 \cos^{-1}\left(\frac{d^2 + r^2 - R^2}{2dr}\right) + R^2 \cos^{-1}\left(\frac{d^2 + R^2 - r^2}{2dR}\right)$$
$$- \frac{1}{2}[(-d + r + R)(d + r - R)(d - r + R)(d + r + R)]^{1/2}. \tag{A.2}$$

$$V = \frac{\pi}{12d}(R + r - d)^2(d^2 + 2dr - 3r^2 + 2dR + 6rR - 3R^2). \tag{A.3}$$

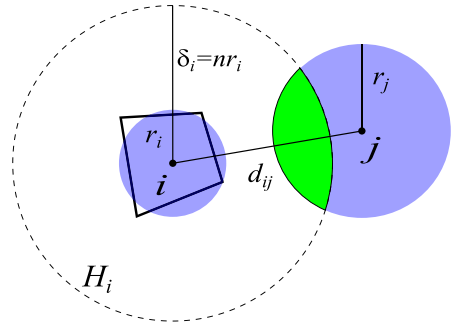

Fig. 23. Volume correction.

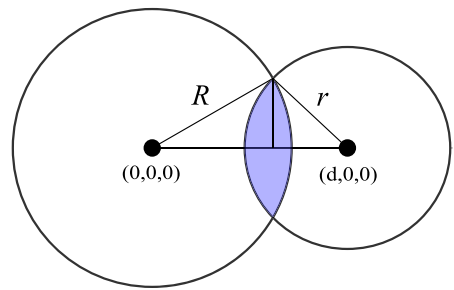

Fig. 24. Circle to circle intersection.